\documentclass[reprint,aps,twocolumn]{revtex4}

\usepackage{graphicx}
\usepackage{dcolumn}
\usepackage{bm}
\usepackage[dvipsnames]{xcolor}

\usepackage{subfigure}
\usepackage{wrapfig}
\usepackage{cancel}
\usepackage{color}
\usepackage{textcomp}
\usepackage{amsmath}
\usepackage{amsfonts}
\usepackage{amssymb}
\usepackage{array}
\usepackage[colorlinks]{hyperref}

\newcommand{\clb}{\color{black}}

\newcommand{\AtBO}{Ag$_2$BiO$_3$}
\newcommand{\BBO}{BaBiO$_3$}

\newcommand{\Tc}{$T_{\text c}$}
\newcommand{\vect}[1]{\boldsymbol{#1}}

\begin{document}

\preprint{APS/123-QED}

\title{Electronic Structure of the Bond Disproportionated Bismuthate \AtBO }

\author{Mohamed Oudah$^{1,2}$}
 \email{mohamed.oudah@ubc.ca}
\author{Minu Kim$^1$}%
\author{Ksenia S. Rabinovich$^1$}%
\author{Kateryna Foyevtsova$^2$}%
\author{Graham McNally$^1$}%
\author{Berkay Kilic$^1$}%
\author{Kathrin K\"uster$^1$}%
\author{Robert Green$^{2,3}$}%
\author{Alexander V. Boris$^1$}%
\author{George Sawatzky$^2$}%
\author{Andreas P. Schnyder$^1$}%
\author{D. A. Bonn$^2$}%
\author{Bernhard Keimer$^1$}%
\author{Hidenori Takagi$^{1,4}$}%
\affiliation{%
$^1$Max Planck Institute for Solid State Research, Heisenbergstrasse 1, 70569 Stuttgart, Germany\\
$^2$Stewart Blusson Quantum Matter Institute, University of British Columbia, Vancouver, British Columbia V6T 1Z4, Canada\\
$^3$Department of Physics \& Engineering Physics, University of Saskatchewan, Saskatoon, Saskatchewan S7N 5E2, Canada\\
$^4$Department of Physics, University of Tokyo, Bunkyo-ku, Hongo 7-3-1, Tokyo 113-0033, Japan
}%

\date{\today}

\begin{abstract}
We present a comprehensive study on the silver bismuthate \AtBO , synthesized under high-pressure high-temperature conditions, which has been the subject of recent theoretical work on topologically complex electronic states. We present X-ray photoelectron spectroscopy results showing two different bismuth states, and X-ray absorption spectroscopy results on the oxygen $K$-edge showing holes in the oxygen bands. These results support a bond disproportionated state with holes on the oxygen atoms for \AtBO . We estimate a band gap of $\sim$1.25~eV for \AtBO\ from optical conductivity measurements, which matches the band gap in density functional calculations of the electronic band structure in the non-symmorphic space group $Pnn2$, which supports two inequivalent Bi sites. In our band structure calculations the disproportionated \AtBO\ is expected to host {\clb Weyl nodal chains, one of which is located   $\sim$0.5~eV below the Fermi level.} Furthermore, we highlight similarities between \AtBO\ and the well-known disproportionated bismuthate \BBO , including breathing phonon modes with similar energy. In both compounds hybridization of Bi-$6s$ and O-$2p$ atomic orbitals is important in shaping the band structure, but in contrast to the Ba-$5p$ in \BBO , the Ag-$4d$ bands in \AtBO\ extend up to the Fermi level.

\end{abstract}

\pacs{Valid PACS appear here}

\maketitle

\section{\label{intro}Introduction}

Transition-metal oxides include many novel classes of materials that host a wide range of structures and electronic properties. The field of condensed matter physics had made a great leap forward with the discovery of high transition temperature superconductivity (\Tc) in the layered copper oxides~\cite{bednorz1986possible} and 
in the {\clb hole-doped} cubic bismuthate (Ba,K)BiO$_3$~\cite{mattheiss1988superconductivity}. {\clb The undoped \BBO\ is a disproportionated insulator.} Also, oxide materials have shown promise for exploring topological properties~\cite{seki2012observation, kargarian2013topological, ishiwata2020emergent}. We identified the silver bismuthates as a promising set of materials for studying highly correlated electron systems and for finding topologically non-trivial oxides. \AtBO\ was recently highlighted in theoretical work with various predictions of topologically protected states~\cite{he2018tunable,fu2018hourglasslike,singh2018topological}. 

The previous experimental work on \AtBO\ has resulted in a number of crystal structures reported in the Inorganic Crystal Structure Database (ICSD), which fall into the space groups $Pnna$, $Pnn2$, and $Pn$~\cite{deibele1999bismuth,oberndorfer2006charge}. $Pnna$ is the oldest structure reported, {\clb which has a single bismuth site} implying a uniform 4+ state for bismuth and as a result equal Bi-O distances. A Bi$^{4+}$ state is unusual, however, and typically a compound with nominal charge of 4+ on bismuth disproportionates. The disproportionation in \BBO\ is reflected in the existence of two distinct sites for Bi, but the nature of this disproportionation is debated. It was emphasized in a following {\clb neutron diffraction study} that \AtBO\ has two distinct bismuth sites and crystalizes in the $Pnn2$ space group~\cite{oberndorfer2006charge}, and the two sites were explained with the disproportionation of Bi$^{4+}$ into Bi$^{3+}$ and Bi$^{5+}$  states. The existence of two distinct bismuth sites in \AtBO\ ($Pnn2$) is similar to \BBO , but these compounds crystallize in different crystal structures, with a combination of corner- and edge-sharing BiO$_6$ octahedra in \AtBO\ and only corner-sharing BiO$_6$ octahedra in {\clb the perovskite \BBO }. By thoroughly characterizing the {\clb disproportionated} state in \AtBO\ and comparison to \BBO , we establish $(M^{1+})_2$BiO$_3$ as a new class of disproportionated bismuthates beyond those forming in the perovskite structure.

{\clb The disproportionation of the Bi site and resulting $Pnn2$ symmetry also have important ramifications for the topological nature of \AtBO .} Theoretical predictions of topological phases in \AtBO\ include an hourglass nodal net semimetal~\cite{fu2018hourglasslike}, a Dirac semimetal~\cite{he2018tunable}, and an hourglass Dirac semimetal~\cite{singh2018topological}. All of these topological predictions were made on \AtBO\ in the $Pnna$ phase with a single bismuth site, where breaking Bi charge order to attain the $Pnna$ symmetry is important to realizing these topologically nontrivial phases. However, we will show that \AtBO\ crystallizes in the $Pnn2$ space group {\clb with two crystallographically distinct bismuth sites} even when synthesized under high-pressure and high-temperature conditions. The topological predictions made for the $Pnna$ phase therefore do not apply to \AtBO\ crystallizing in the $Pnn2$ phase. {\clb Rather, we predict Weyl states in the $Pnn2$ phase}.

\begin{figure*}
\includegraphics[width=16cm]{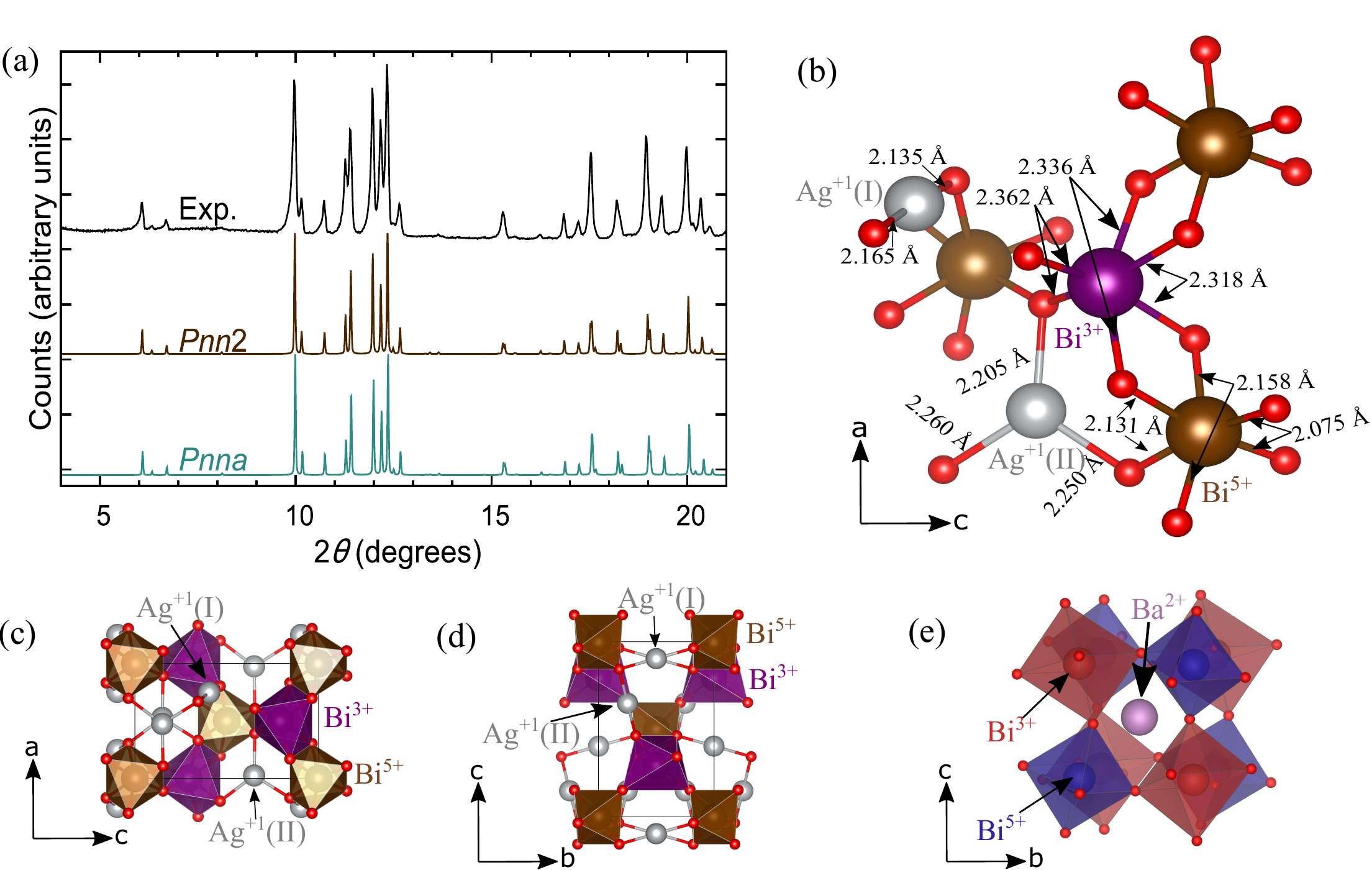}
\caption{(a) The powder X-ray diffraction pattern measured at room temperature of \AtBO\ (top) and the calculated pattern for \AtBO\ in the $Pnn2$ phase (middle) and the $Pnna$ phase (bottom). (b) Part of the unit cell of \AtBO\ in the $Pnn2$ phase with ``Bi$^{3+}$'' (purple) and ``Bi$^{5+}$'' (brown) highlighting the different oxygen metal distances~\cite{oberndorfer2006charge}. (c) Unit cell of \AtBO\ in the $Pnn2$ phase along the $ac$-plane showing the shared edges of the octahedra along the $a$ direction. (d) Unit cell of \AtBO\ in the $Pnn2$ phase along the $bc$-plane. (e) Unit cell of \BBO\ showing the corner-sharing octahedra and ``Bi$^{3+}$'' (red) and ``Bi$^{5+}$''(blue).}
\label{Fig1_XRD}
\end{figure*}

In this paper, we characterize \AtBO\ samples synthesized at 6~GPa using {\clb powder X-ray diffraction (XRD)} and various spectroscopic techniques. {\clb We present evidence from X-ray photoelectron Spectroscopy (XPS) supporting the disproportionation of bismuth into two sites in \AtBO , and present evidence for a bulk band gap, a consequence of the disproportionation, seen in an optical conductivity measurement.} The optical gap is consistent with density functional theory (DFT) calculations of the electronic band structure in the $Pnn2$ (disproportionated) phase of this material. {\clb Our calculations, on the experimentally realized $Pnn2$ phase, further predict Weyl nodal chains (i.e., chains of connected loop degeneracies in momentum space) at various energies in this phase, one being $\sim$0.5~eV below the Fermi level.} We provide evidence for oxygen holes using X-ray absorption spectroscopy (XAS), and considering the XPS results we propose that \AtBO\ is bond disproportionated rather than charge disproportionated.
{\clb We then} highlight the extent of similarities between \AtBO\ and \BBO ; similar breathing mode peaks are observed in Raman spectra and we identify similarities in the band structure of these compounds when considering the oxygen molecular orbitals in these disproportionated bismuthates. Finally, we present new evidence from Raman spectra and XRD for the previously reported structural transition to a $Pn$ phase at low temperature~\cite{oberndorfer2006charge}.

\section{Experimental Details}

\AtBO\ samples were synthesized with a high-pressure, high-temperature method utilizing a belt-press. Bi$_2$O$_3$ and AgO powders were well mixed in a mortar and pestle, then loaded into a gold capsule with a drop of KOH to speed up the reaction. The capsule was then pressed to 6~GPa and heated at 730~K for 1~hour before quenching and retrieving the polycrystalline product. A finely ground sample was loaded into a 0.3~mm capillary, and powder XRD patterns of \AtBO\ were collected while rotating the capillary using Ag radiation (0.5594~\AA). {\clb The samples were cooled down to ${\sim}200~$K for low temperature XRD.}

\begin{figure*}
\includegraphics[width=16cm]{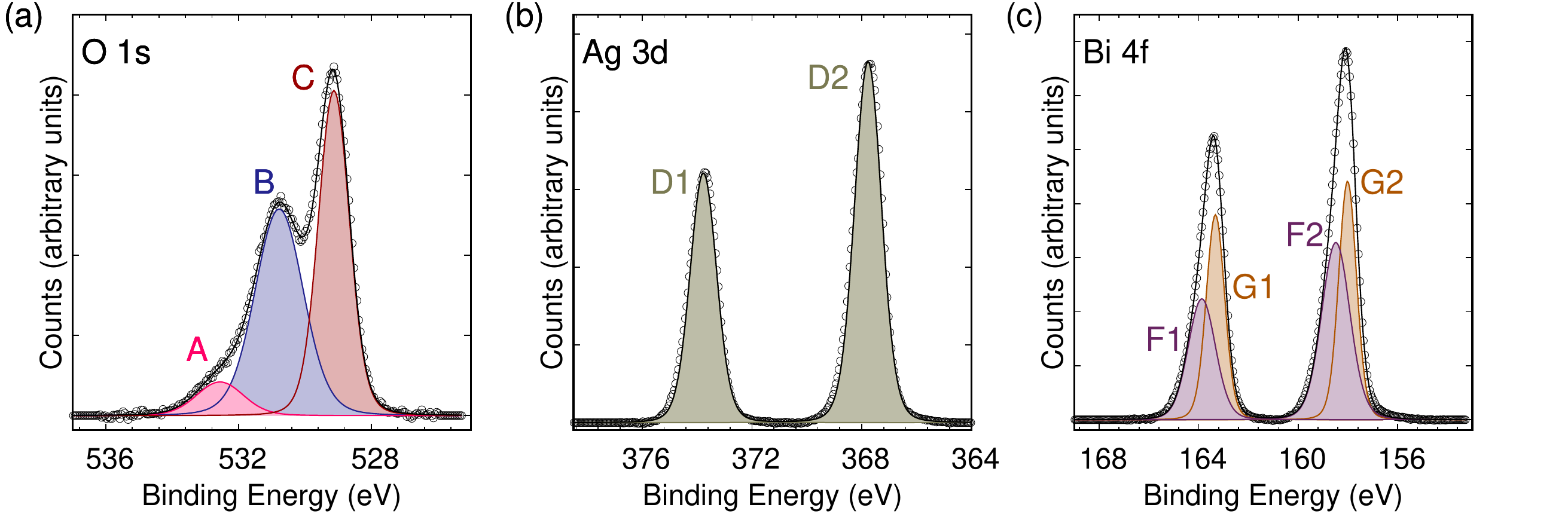}
\caption{X-ray photoelectron spectra of O, Ag, and Bi in \AtBO\ are shown in (a), (b), and (c). In (a) the three peaks A, B, and C are needed to fit the measured spectra for the O-$1s$ states. In (b) the splitting between peaks D1 and D2 is consistent with that expected due to spin-orbit-coupling (SOC) for Ag-$3d$ states. In (c) the splitting between F1 and F2 is consistent with that expected due to SOC for Bi-$4f$ states, and the same can be said for the split between G1 and G2. Due to the asymmetry of the measured spectra, four curves are needed to fit the Bi-$4f$ states. This leads us to conclude the existence of two different Bi states in \AtBO , both close in energy to Bi$^{3+}$~\cite{hegde1989electronic}.}
\label{Fig2_XPS}
\end{figure*}

X-ray absorption spectra of the oxygen $K$-edges were recorded in partial fluorescence yield (PFY) mode at the Spherical Grating Monochromator (SGM) beamline at the Canadian Light Source.  Spectra in PFY mode are collected using energy resolving silicon drift detectors, which allow selection of only the oxygen $K$-edge fluorescence. The average of data from four separate silicon drift detectors was used for the PFY spectra. Spectra are calibrated in energy by using a TiO$_2$ reference sample~\cite{quackenbush2013nature}. Spectra were then normalized by the incident beam intensity, and normalized according to post-edge intensity after subtracting a linear background.  The \AtBO\ spectra were collected under ultrahigh vacuum at room temperature from a pressed pellet attached to carbon tape, with the pellet surface at an angle of 45° to the incident beam.
The XPS data were collected using a commercial Kratos AXIS Ultra spectrometer and a monochromatized Al $K_\alpha$ source (photon energy: 1486.6~eV). The base pressure during XPS was in the low $10^{-10}$~mbar range. The spectra were collected using an analyzer pass energy of 20~eV. XPS spectra were analyzed using the CasaXPS software. The charge-neutralizer was used and the C 1s was set to 284.8 eV for binding energy calibration~\cite{swift1982adventitious,barr1995nature}. To fit the peaks, a combination of multiple Gaussian-Lorentzian mixture functions and a Shirley background was used. 
The binding energy separation and the area ratio of the doublets were not constrained but the results are within experimental error of the tabulated values~\cite{moulder1992handbook}. The sum of all the fitted functions to XPS peaks is shown with a black curve in Fig.~\ref{Fig2_XPS}.

Optical spectroscopic ellipsometry measurements were performed using a variable-angle spectroscopic ellipsometer (VASE, Woollam) with photon energy in the range of 0.7–-6.5~eV at incident angles of 70$^{\circ}$. The real and imaginary parts of the dielectric function can be obtained accurately from the ellipsometric angles $\psi (\omega)$ and $\Delta (\omega)$ extracted from the optical spectra, without the need of a Kramers-Kronig analysis.

The Raman data were collected in backscattering geometry using a Dilor XY-triple grating Raman spectrometer equipped with a charge coupled device camera as detector. With power of 0.2~mW at 632~nm it was possible to measure Raman spectra on polished surfaces of the \AtBO\ samples. Low temperature Raman measurements were performed between 10-300~K. The resolution of our spectrometers for this experiment was about 3~cm$^{-1}$.

\section{Computational Details}

The band structure calculations for \AtBO\  presented in Fig.~\ref{Optical}~(d) were obtained using the Vienna ab initio Simulation Package (VASP 5.4) with the generalized gradient approximation (GGA) of Perdew-Burke-Ernzerhof (PBE)
type exchange-correlation potential~\cite{perdew1996generalized}. The cut-off energy of plane wave basis set was set to be the ENMAX value in the pseudo-potential plus $30\%$. A-centered $3\times3\times2$ Monkhorst-Pack grid was used for the self-consistent field (SCF) calculations.

The electronic structure calculations shown in Figs.~\ref{BBOband}~{\color{blue}(a)-(d)}
were performed using density functional theory (DFT) with the full-potential linearized-augmented-plane-wave code WIEN2k~\cite{blaha2001wien2k}.
Exchange and correlation effects were treated within the local density approximation (LDA)~\cite{Perdew92}.
For \AtBO, we use the $Pnn2$ unit cell with
the lattice constants $a=5.983$~\AA,
$b=6.324$~\AA,
and $c=9.576$~\AA, as determined in Ref.~\cite{oberndorfer2006charge}.
For \BBO, we use the $P2_1/n$ unit cell with
the lattice constants $a=6.174$~\AA,
$b=6.125$~\AA,
and $c=8.652$~\AA, as determined in Ref.~\cite{Kennedy06}.
For both bismuthates,
a $6\times6\times4$ $\vect{k}$-point grid was used for the Brillouin-zone integration.
Projections onto oxygen molecular orbitals
were performed with a modified version of WIEN2k, following
the procedure outlined in Ref.~\cite{Foyevtsova19}.

\clearpage

\section{Results and Discussion}
\subsection{\label{spectra}Analysis of XRD, XPS, and XAS}

{\clb We used XRD, XPS and XAS measurements to demonstrate that \AtBO\ in the $Pnn2$ phase is a bond-disproportionated bismuthate.} As it is difficult to distinguish the $Pnna$ structure from the $Pnn2$ structure with only PXRD data, as demonstrated in Fig.~\ref{Fig1_XRD}, we will start by accepting the reported $Pnn2$ phase reported and assess the consistency of various properties we measure with this structure. {\clb The lower scattering power of oxygen in X-ray diffraction and natural twinning of \AtBO\ crystals makes it difficult to discern the $Pnn2$ from the $Pnna$ structure. A previous powder neutron diffraction study, which is more sensitive to diffraction from oxygen than X-rays, on \AtBO\ resulted in a precise refinement of O atomic positions and a structure where Bi has two distinct sites~\cite{oberndorfer2006charge}. We compare the simulated powder XRD and neutron diffraction patterns in Sup.~Fig.~1 to demonstrate how the $Pnn2$ phase can be identified in neutron diffraction~\cite{SupMat}.}

In the $Pnn2$ phase, \AtBO\ has two distinct bismuth sites resulting in two BiO$_6$ octahedra with shorter and longer Bi-O bonds. It should be emphasized that the Bi-O distances vary within each of the two types of octahedra in \AtBO . These Bi-O$_6$ octahedra are edge-sharing along the $a$ direction, but only share corners along the $b$ and $c$ directions. The O-O distances at the shared edges of these octahedra are shorter than the distances between the other oxygen atoms, which may correspond to oxygen dimerization at the shared edges. The Ag atoms are sitting between the BiO$_6$ octahedra channels and are bonded to 3 oxygen atoms each. The shortest Ag-O distances for both Ag sites are 2.1464 and 2.2065~\AA , which are expected for Ag$^{1+}$ typically seen in oxides.
\begin{figure*}
\includegraphics[width=16cm]{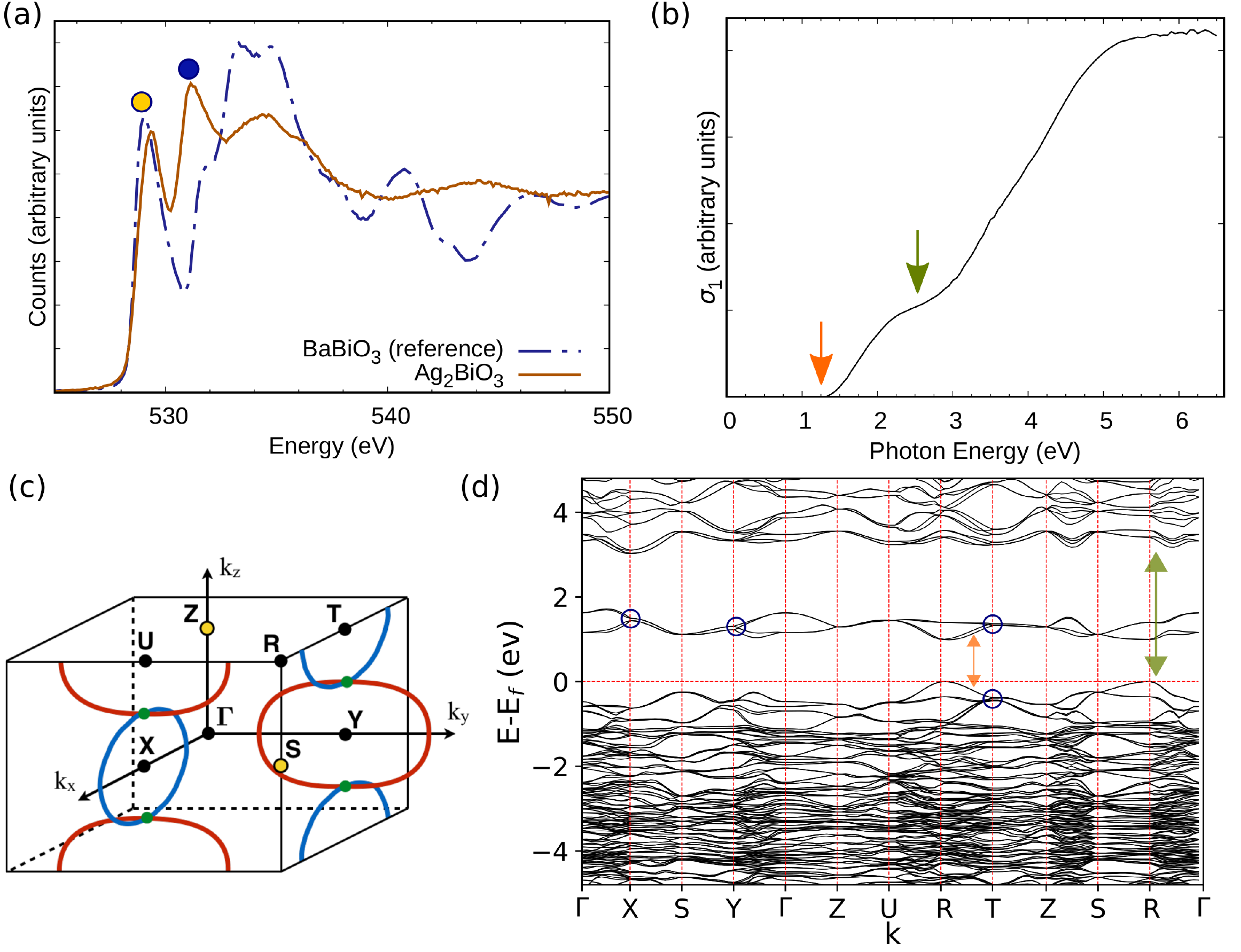}
\caption{(a) The X-ray absorption spectra at the Oxygen $K$-edge of \AtBO\ and a reference sample of \BBO . 
(b) Optical spectroscopy data on \AtBO\ extracted from ellipsometry measurement. 
(c) {\clb Brillouin zone of \AtBO with the high-symmetry points indicated by black dots. The nodal chains are schematically
illustrated by the blue and red lines. At the Z and S points (yellow) there are fourfold degeneracies, formed by two Weyl points with opposite chirality.}
(d) Band structure of \AtBO\ in the $Pnn2$ phase using DFT calculation with hybrid functionals.}
\label{Optical}
\end{figure*}

The X-ray photoelectron spectra for Ag, Bi, and O in \AtBO\ are presented in Figs.~\ref{Fig2_XPS}(a)-(c). {\clb As expected, only} one state is observed for Ag, with peak centers at 373.8~eV (peak D1 in Fig.~\ref{Fig2_XPS}(b)) and 367.8~eV (peak D2 in Fig.~\ref{Fig2_XPS}(b)) {\clb which} match the transition seen in Ag$^{1+}$ in other oxides and chalcogenides~\cite{moulder1992handbook}. The splitting here is expected for the $d$-states due to spin-orbit-coupling (SOC). {\clb The XPS of O shows three peaks at 529.1~eV (C), 530.8~eV (B), and 532.6~eV (A). 
As for the minor peak at 532.6~eV (peak A in Fig.~\ref{Fig2_XPS}(a)), it was attributed to surface contaminants in other bismuthates~\cite{balandeh2017experimental}, and such contaminants may be present on the surface of our samples.
The binding energy of the main peak for O $1s$ at 529.1~eV (peak C in Fig.~\ref{Fig2_XPS}(a)) is comparable to the binding energy of O~$1s$ in \BBO ~\cite{balandeh2017experimental}. 
XPS shows evidence for KOH on the surface of the sample as evidenced by an observed K peak (shown in Sup.~Fig.~2(a)); we cannot rule out that some of the intensity at at 530.8~eV originates from KOH~\cite{favaro2016unravelling}(peak B in Fig.~\ref{Fig2_XPS}(a)). Furthermore, contamination like carbonates, which were also observed in the C~$1s$ spectrum (see Sup.~Fig.~2(a)) might also contribute the peak B. However, analyzing the relative amount of potassium and carbonate contaminations peak B cannot be purely due to those contaminations, see Sup.~Table~I. 
Considering, that the area of the fitted curve at 530.8~eV is comparable to the area of the main peak at 529.1~eV (Table I in SI), we expect the signal to originate from a similar amount of oxygen atoms on the surface of the \AtBO\ samples. Looking at the different O-O and Bi-O distance for oxygen atoms at the shared edges and corners of the BiO$_6$ octahedra, we propose that the peaks at 530.8~eV and 529.1~eV may originate from different states in the oxygen atoms at the shared corners and the shared edges of the BiO$_6$ octahedra. }

\begin{figure*}
\includegraphics[width=16cm]{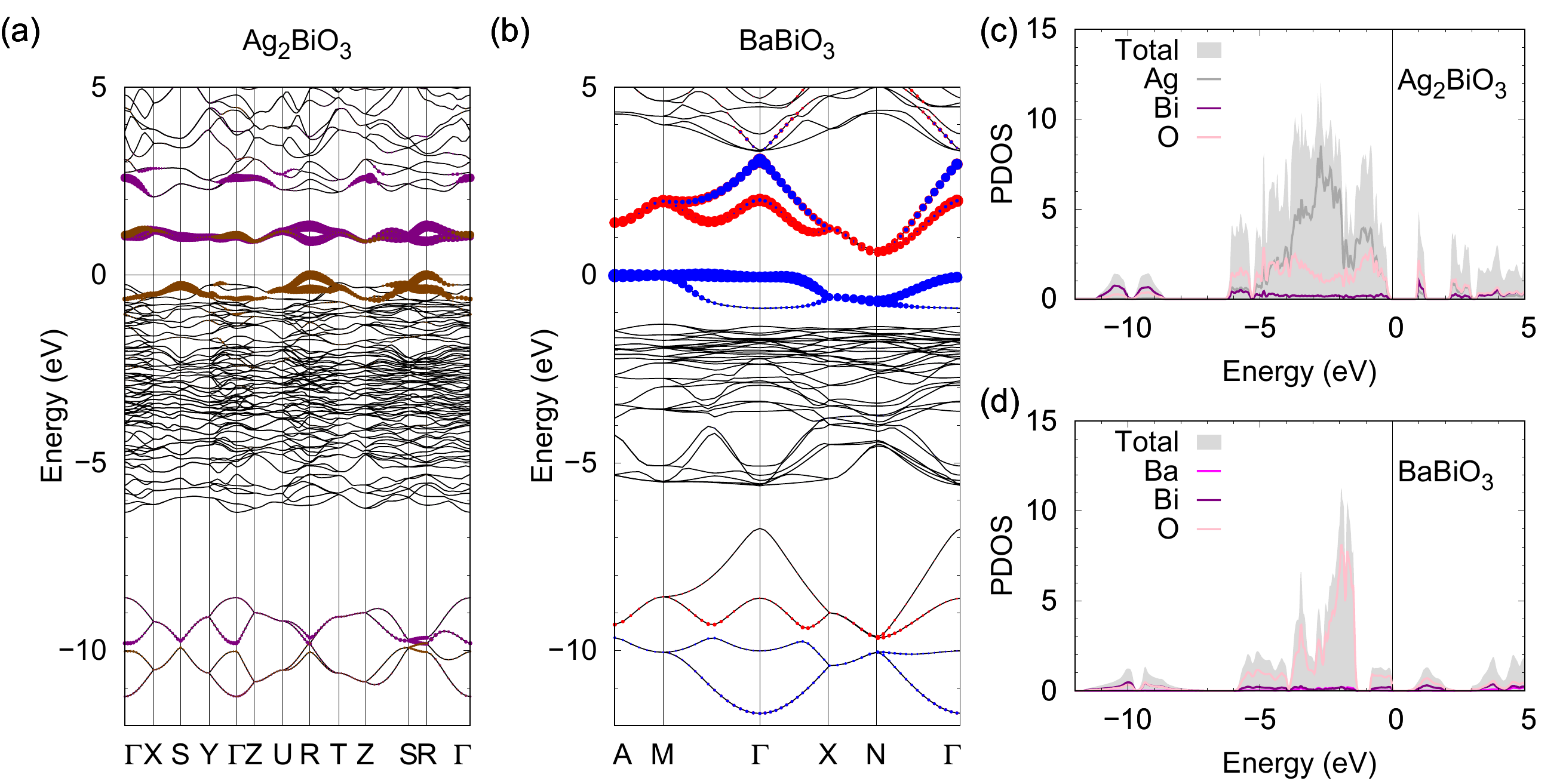}
\caption{(a) The DFT (LDA) band structures of \AtBO\ in the $Pnn2$ phase.
The contribution from the O-$a_{1g}$ molecular orbital of the small (purple) and large (brown) octahedra are represented with fat bands. The colors for octahedra are matching the \AtBO\ unit cell presented in Fig.~\ref{Fig1_XRD}(c). 
(b) The band structures of \BBO\ in the $P2_1/n$ phase. The contribution from the O-$a_{1g}$ molecular orbital of the small (red) and large (blue) octahedra are represented with fat bands. The colors for octahedra are matching the \BBO\ unit cell presented in Fig.~\ref{Fig1_XRD}(e). 
(c) Partial Density of States (PDOS) of \AtBO , showing the contribution from Ag, Bi, and O.
(d) PDOS of \BBO , showing the contribution of Ba, Bi, and O.}
\label{BBOband}
\end{figure*}

Assigning distinct oxygen atoms as the origin for the different XPS peaks, the peak at 529.1~eV may originate from the dimerized oxygen at the shared edges of the octahedra with a bond length of 2.6588~\AA , while the contribution to the peak at 530.8~eV from \AtBO\ likely comes from the corner shared oxygens. Another possibility is that the different oxygen states at 529.1 and 530.8~eV arise from the interaction of the oxygen atoms with silver atoms at different distances, where the oxygen atoms at the shared corners of the octahedra are further away from the nearest silver atoms, while the oxygen atoms at the shared edges are closer to silver atoms. {\clb If all oxygen atoms in \AtBO\ have equivalent Bi-O interaction, then the electronic origin for the split peaks at 529.1 and 530.8~eV depends on the extent of the O~$1s$ core-hole potential and the Bi~$6s$/O~$2p$ hybridization~\cite{balandeh2017experimental}}.

The Bi $4f$ spectra, shown in Fig.~\ref{Fig2_XPS}(c), support the presence of two Bi states in this compound, {\clb and} these states have similar binding energy (158.0~eV for peak G2 and
158.5~eV for peak F2). {\clb The ratio of the areas of the two peaks is nearly 1:1, suggesting that the two different Bi states are due to the two distinct Bi atoms in the unit cell. It might be possible that one of the Bi states is due to surface contaminations like hydroxides or carbonates. However, this would imply that the 1:1 ratio of the two Bi states observed is mere coincidence. Furthermore, if one of the components would be due to contamination the amount of Bi would be too little compared to the Ag amount (see Table I in SI). For now, we consider the two Bi peaks observed as originating from \AtBO .
Comparing the measured binding energies with those for Bi$^{5+}$ like in NaBiO$_3$~\cite{kulkarni1990state} and Bi$^{3+}$ like in Bi$_2$O$_3$~\cite{hegde1989electronic} we realize that both are close to the value for Bi$^{3+}$. This might hint that the system is bond disproportionated rather than charge disproportionated. Please note that the Bi~$4f$ binding energies are similar to the one observed in \BBO ~\cite{plumb2016momentum} but the Bi~$4f$ spectrum for \AtBO\ shows a clearer asymmetry. }


{\clb The XPS spectra thus indicate two different Bi sites, consistent with a disproprotionated state, and optical conductivity evidence for a band gap (Section~\ref{Topology}) support this interpretation. However, whether \AtBO\ is charge or bond disproportionated remains unclear without examining the XAS data. XAS results at the oxygen $K$-edge, presented in Fig.~\ref{Optical}(a), show a sharp pre-peak at 529~eV, which originates from a high hole density in the oxygen $p$ orbitals. This is similar to the results on \BBO ~\cite{balandeh2017experimental}.} The higher energy region at the $K$-edge is clearly different from the \BBO\ spectrum, reflecting the different oxygen environment in \AtBO . 
{\clb A second peak is seen in the XAS results of \AtBO , and the two XAS peaks (marked with yellow and blue circles in Fig.~\ref{Optical}(a)) are $\sim$2~eV apart. This energy difference is similar to peaks B and C in the oxygen XPS spectra.} 

The combination of corner and edge sharing octahedra in \AtBO\ can contribute to the different XAS spectra above 530~eV. Another peak at 531~eV is seen for \AtBO , and the origin of this peak is the higher energy states of O in \AtBO. If we compare Figs.~\ref{BBOband}(c) and (d), we see a peak in the oxygen partial density of states at 2 eV in \AtBO ~that is much less intense than in \BBO . In the ionic-limit, we have Ag$^{1+}$--$4d^{10}$, Bi$^{3+}$--$6s^2$, and O$^{2-}$--$2p^6$ ions making up \AtBO , where the BiO$_6$ octahedron are either made up of Bi$^{3+}$($6s^2$)--O bonds with longer bonds or Bi$^{3+}$($6s^2\underline{L}^2$)--O with shorter bonds. The ligand hole pairs ($\underline{L}^2$) condense on these smaller octahedra, where the O atoms have equal charge on them. The disproportionation leading to the small and large octahedra in this case is described as bond-disproportionation.

One is tempted to conclude a similar origin for both of these sets of peaks, such as the difference between oxygen atoms at the shared edges and corners of the BiO$_6$ octahedra. However, the area under the curve for the XPS peaks is equivalent for B and C. {\clb Reconciling the XAS and XPS results, if the oxygen atoms are equivalent in terms of interaction with Bi and the split in the O~$1s$ XPS is electronic in nature, then an oxygen ligand hole will be distributed over 6 oxygen atoms in the BiO$_6$ octahedron.} In the case of O~$1s$, the XPS peak splitting originates from oxygen atoms at different distances to bismuth, which suggests an oxygen hole may be distributed over 4 oxygen atoms at the shared edges of the BiO$_6$ octahedra. The oxygen hole ligand distribution in the octahedra of \AtBO\ is discussed below in section ~\ref{BBO}.


\subsection{\label{Topology} Band Gap and Topology\\ in Disproportionated \AtBO}

The band structure of \AtBO\ in the $Pnn2$ phase resulting from DFT calculations is presented in Fig.~\ref{Optical}(d), where we have a band gap of $\sim1.25$~eV resulting from hybridization of O $2p$ bands and Bi $6s$ band, with anti-bonding character. The second set of bands above the Fermi level ($E_{\rm{F}}$) is located $\sim3$~eV above $E_{\rm{F}}$ and corresponds to the unfilled Bi~$6p$ bands. The band gap from optical conductivity measurement is around {\clb $\sim 1.25$~eV (Fig.~\ref{Optical}(b)), transition between bands marked with an orange arrow}, which corresponds well with transitions expected in our calculated band structure, marked with an orange arrow in {\clb Fig.~\ref{Optical}(d)}. {\clb The real and imaginary parts of the dielectric constant of \AtBO\ are shown in Sup.~Fig.~3.} The upturn starting at $\sim2.5$~eV in the optical conductivity corresponds to transitions from bands below $E_{\rm{F}}$ to the empty Bi~$6p$ bands in \AtBO , which are marked with green arrows in {\clb Figs.~\ref{Optical}(b) and (d)}. The consistency between the features seen in our optical conductivity measurement and the expected transitions in our calculated band structure provide further evidence for \AtBO\ being in the disproportionated $Pnn2$ phase. In the $Pnna$ phase \AtBO\ is a semimetal and expected to have a nonzero density of state at the Fermi level~\cite{he2018tunable}, so no gap in the optical conductivity. 

{\clb We emphasize that the non-symmorphic space group $Pnn2$ (Space Group 34) of  \AtBO\  leads to symmetry-enforced nodal loops in all bands, i.e., twofold Weyl crossings along 1D loops. These nodal loops are protected by the glide mirror symmetries \{m$_{010}$$|$$\tfrac{1}{2}$ $\tfrac{1}{2}$ $\tfrac{1}{2}$\} and \{m$_{100}$$|$$\tfrac{1}{2}$ $\tfrac{1}{2}$ $\tfrac{1}{2}$\} and touch each other on the intersection lines of the mirror planes, thereby forming ``nodal chains" \cite{bzdusek_nature,chan2019symmetry,hirschmann_arXiv}, see Fig.~\ref{Optical}(c). These Weyl nodal lines carry
a nontrivial $\pi$-Berry phase, leading to drumhead surface states.  
The Weyl nodal-chain feature is not removable by perturbations as long as the glide mirror symmetries are preserved. 
In our band calculations of Fig.~\ref{Optical}(d) these Weyl nodal chains are seen in various bands as crossing points close
to high-symmetry points (see, e.g, crossings marked by blue circles close to the X, Y, and T points).
To access the anomalous magnetotransport properties of the nodal chains~\cite{bzdusek_nature},
we may be tempted to use chemical doping to shift the Fermi level and approach these band crossings. However,
if the chemical doping changes the $Pnn2$ symmetry, then these topological features may be annihilated.
In addition, we note that at the S and Z points the bands
form fourfold degeneracies, composed of two Weyl points with opposite chirality (yellow dots in Fig.~\ref{Optical}(c)).}

Including hybrid functionals in our calculations results in an increase in the band gap for \AtBO\ in comparison with calculations in the literature~\cite{he2018tunable} with a direct gap of $\sim~1.0$~eV at the $R$ point in the band structure. However, this value remains slightly lower than the experimentally estimated value of $\sim~1.25$~eV from our optical conductivity measurement. This may be limited by the optically active transitions in \AtBO\ that do not correspond to the lowest band gap in our band structure, or the inadequacy of the band calculations in providing accurate band gap values even with hybrid functionals. \AtBO\ in the $Pnn2$ is predicted to be a good photovoltaic material~\cite{fabini2019candidate}, and our results clarifying the band gap experimentally are an important step forward. The $\sim~1.25$~eV band gap in \AtBO\ falls within 1.1-1.45~eV range where maximum efficiency is expected for single junction solar cells~\cite{kirchartz2018makes}. {\clb  XPS results, shown in Sup.~Fig.~2(b), suggest the valence band is $\sim 0.9$~eV below the Fermi level, and this shows promise for introducing in-gap-states in \AtBO .} In future studies, we will need to examine the stability of \AtBO\ when exposed to different radiation to assess its potential in photovoltaic applications. 

\subsection{\label{BBO}Comparing \AtBO\ and \BBO}

Next, we would like to highlight the extent of similarity and some of the differences between \AtBO\ and \BBO.
Both materials have a bulk band gap due to disproportionation of ``Bi$^{4+}$",
but the octahedra are only corner sharing in \BBO\ while the octahedra are sharing edges and corners in \AtBO . The nominal Bi$^{4+}$ oxidation state that leads to the disproportionated charge/bonds in these compounds is achieved through a different number of atoms. 
While having one Bi atom and three oxygen atoms, it takes two monovalent Ag$^{1+}$ ions or one divalent Ba$^{2+}$ ion to achieve the same average formal charge of 4+ for Bi. Thus, (Ag$^{1+}$)$_2$(Bi$^{4+}$)(O$^{2-}$)$_3$ and (Ba$^{2+}$)(Bi$^{4+}$)(O$^{2-}$)$_3$ have the same unstable average formal charge for Bi in different crystal structures, and lead to disproportionation in both \AtBO\ and \BBO .
However, similar to \BBO, there is a significant amount
of oxygen character in the empty conduction states of \AtBO\
above the Fermi level. To be more specific, these states
are formed by oxygen molecular orbitals with $a_{1g}$ symmetry
centered on the \emph{small} BiO$_6$ octahedra. In Fig.~\ref{BBOband}~(a),
they are highlighted as purple fat bands for \AtBO,
while in Fig.~\ref{BBOband}~(b),
they are highlighted as red fat bands for \BBO.
In the previous theoretical studies
on \BBO\cite{Foyevtsova15,khazraie2018oxygen},
this situation was referred to as a bi-polaronic condensation
of oxygen holes. {\clb For \AtBO\ we show in Sup.~Fig.~4 the orbital contribution from Bi-$s$ and $p$ in the small and large O$_6$ octahedral cages, as well as the O-2$p$. The significant contribution of O-2$p$ above the Fermi level is consistent with high density of oxygen holes as evidenced by the pre-peak in XAS (Fig.~\ref{Optical}(a)).}

\begin{figure*}
\includegraphics[width=16cm]{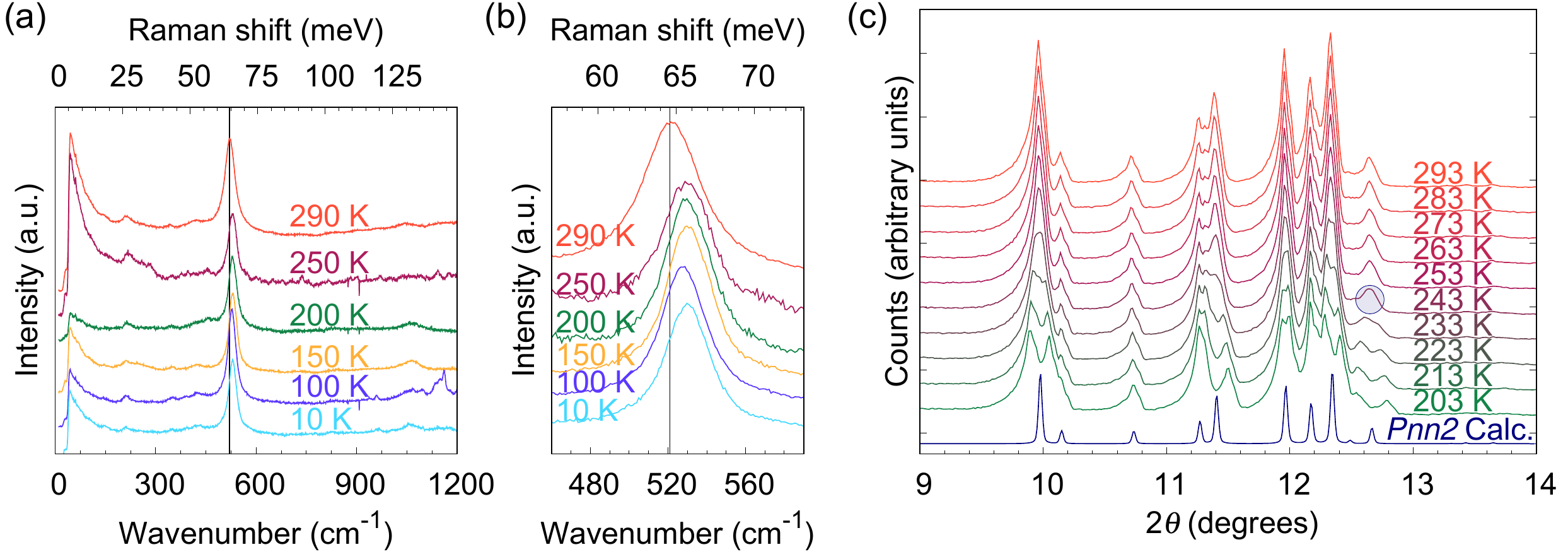}
\caption{(a) Raman spectra of \AtBO\ at different $T$, between 290--10~K. (b) The same Raman spectra in (a), emphasizing the peak at 520~cm$^{-1}$, characteristic of the breathing mode in disproportionated bismuthates. (c) Powder-XRD of \AtBO\ collected at different $T$, between 293--203~K. The calculated XRD pattern for \AtBO\ in the $Pnn2$ phase is plotted at the bottom.}
\label{Raman}
\end{figure*}

As one can also see in Fig.~\ref{BBOband},
the valence state within $\sim$1~eV range below the Fermi level
has, in turn, a strong character of O$\textendash a_{1g}$
molecular orbitals centered on the \emph{large} BiO$_6$ octahedra
(brown fat bands for \AtBO\ and blue fat bands for \BBO).
Together, these two sets of oxygen molecular orbital states
around the Fermi level constitute the anti-bonding
manifold resulting from the strong hybridization
between the Bi$\textendash 6s$ and O$\textendash a_{1g}$ orbitals.
The latter appears to be as strong in \AtBO\ as it was found
to be in \BBO,
which is evident from the fact that in both systems the bonding
combination is pushed down as far as -10~eV.
A new feature of the \AtBO\ band-structure is
the presence of Ag$\textendash 4d$ bands
in the same energy region where the non-bonding
oxygen bands are located, from $\sim$-7~eV up to the Fermi level {\clb (Shown in Sup.~Fig.~4(a))}.
It was demonstrated in \BBO\ that {\clb the} octahedral angle between Bi-O$_6$ octahedra effects the Bi-O hybridization~\cite{khazraie2018oxygen}. Band calculations in \BBO\ with a 0$^{\circ}$ and 16$^{\circ}$ octahedral tilting show bands with narrower energy dispersion with increased tilt angle.
The conduction band in \AtBO\ is considerably
narrower compared to that in \BBO, which is probably
a result of weaker Bi-O hybridization due to the geometrical arrangement
of the BiO$_6$ octahedra in \AtBO . In \AtBO , the edge-sharing Bi-O$_6$ octahedra along the $a$ direction, and the high angle of 36$^{\circ}$ between the Bi-O-Bi at the shared corners along the $b$ and $c$ directions contribute to reduced hybrization. This qualitative statement about the hybridization should be confirmed with DFT calculations in future studies. The Ag might play a role in reducing the band gap in \AtBO , as seen in the Ag-containing pyrochlore oxides~\cite{mizoguchi2004electronic}. This narrowing effect is not clear when we compare \AtBO\ with \BBO\ due to the different crystal structure, but replacing the Ag with another monovalent atom, such as Na$_2$BiO$_3$, can help us to study the effect of Ag. The contribution from Ag bands near the Fermi level is shown in the partial density of states (PDOS) of \AtBO\ presented in Fig.~\ref{BBOband}(c). In comparison, the PDOS of \BBO , shown in Fig.~\ref{BBOband}(d), has significant contribution from only Bi and O near the Fermi level and no significant contribution from the Ba.

While in \BBO\ we have an indirect band gap as shown in Fig.~\ref{BBOband}~(b),  in \AtBO\ the band gap is direct as shown in Fig~\ref{Optical}(d) and Fig.~\ref{BBOband}~(a). The role of the  Ag$\textendash 4d$ bands that extend over a wide energy range below the Fermi level is not clear for now. Identifying the effect of edge sharing BiO$_6$ octahedra, crystal symmetry, and Ag bands on the band structure of \AtBO\ will be an important part of future studies. Future work should include replacing Ag with other monovalent metal $M^{+1}$ atoms, varying tilting angle of Bi-O$_6$ octahedra, and studying the lower symmetry $Pn$ phase.

The Raman spectra of \AtBO\ have a strong phonon mode at 520~cm$^{-1}$, shown in Fig.~\ref{Raman}, which is similar to the phonon mode observed in pristine \BBO\ at 570~cm$^{-1}$~\cite{tajima1992raman}. In the case of \BBO , the phonon is attributed to the breathing mode originating from local distortion in the bond disproportionated \BBO , so it is likely that a similar energy phonon originates from a similar breathing mode accompanying the disproprotionation in \AtBO . The slight difference in energies between the phonon modes in \AtBO\ and \BBO\ is due to the different arrangement of the BiO$_6$ octahedra in the two compounds and the slightly different Bi-O distances. A higher energy phonon mode originating from higher harmonic phonon modes can be seen in the Raman spectra of \AtBO\ at 1100~cm$^{-1}$, which is similar in value to phonons in the spectra of pristine \BBO ~\cite{tajima1992raman}. 



\subsection{\label{PhaseTrans}Transition to $Pn$ Phase in \AtBO}

We have closely measured the phonon mode at 520~cm$^{-1}$ as a function of temperature between 290~K and 10~K, and have observed a shift in the phonon peak position cooling down from 290~K to 250~K, but no further shift in the phonon mode cooling down to 10~K. We have performed powder XRD measurements on \AtBO\ at low $T$ and observe a splitting in some peaks at 240~K that corresponds to the monoclinic  distortion previously reported~\cite{oberndorfer2006charge}. This distortion leads to reduction of symmetry of \AtBO\ from $Pnn2$ to $Pn$. 

As explained in Sec.~\ref{spectra}, it is difficult to distinguish whether \AtBO\ crystallizes in the $Pnn2$ or the $Pnna$ phase from XRD alone, and neutron diffraction studies were required to clarify the $Pnn2$ phase at room temperature~\cite{oberndorfer2006charge}. The band gap and splitting in the Bi XPS states measured on our samples suggest that \AtBO\ is indeed in the $Pnn2$ phase. However, the $Pn$ phase of \AtBO\ is distinct from the higher symmetry phases and can be identified with XRD. {\clb The splitting of the (202) peak at $\sim12.6^{\circ}$ for temperatures $<243$~K is highlighted with a blue circle in Fig.~\ref{Raman}(c), which results from the reduced symmetry of \AtBO\ $Pnn2$ to $Pn$.} 

It is likely that the shift in the phonon mode observed in Raman measurements at 250~K is related to the phase transition observed in powder XRD. The transition observed in our samples is consistent with the previously reported phase transition in neutron diffraction measurements on \AtBO ~\cite{oberndorfer2006charge}. 
In future studies, we hope to investigate the effect of different synthesis conditions on the oxygen content in \AtBO , and whether this influences the distortion in these samples at low $T$.

Measurements that can be used to probe the band structure of \AtBO\ such as scanning tunneling spectroscopy or angle-resolved photoemission spectroscopy are typically performed at low temperature. Any properties measured on \AtBO\ below $\sim250$~K are representative of the material in the $Pn$ phase, and the band structure of this phase will be the relevant one for those measurements. {\clb Luckily, the $Pn$ (Space Group 7) symmetry falls into the non-symmorphic space groups where a single glide mirror
symmetry is expected to lead to symmetry-enforced Weyl nodal lines~\cite{bzdusek_nature,chan2019symmetry,hirschmann_arXiv}, albeit not a nodal chain as in the $Pnn2$ phase.}

The disproportionation of \AtBO\ is still prevalent in the $Pn$ phase, which would lead to a band gap much like the $Pnn2$ phase. Repeating the XPS and XAS measurements at temperatures where \AtBO\ is in the $Pn$ phase will help reveal whether this transition affects the Bi states or the oxygen holes. 


\section{Conclusions}

We have demonstrated that despite the use of a high pressure of  6~GPa, \AtBO\ samples crystallize in a structure consistent with the $Pnn2$ phase as evidenced by the XPS data for Bi and the measured band gap. The band gap observed in our study is consistent with our band calculations, and we demonstrate that \AtBO\ even in the $Pnn2$ phase is topologically non-trivial and contains {\clb Weyl-nodal chains}   in its band structure ($\sim$0.5~eV below the Fermi level). We observe a phase transition in \AtBO\ at low temperatures associated with distortion from the $Pnn2$ phase to the $Pn$ phase from our XRD pattern and Raman spectra. Even in the $Pn$ phase \AtBO\ falls into a non-symmorphic space group and Weyl nodal-line features are expected. 

The XAS spectra of \AtBO\ suggests the presence of oxygen holes much like \BBO , which is consistent with \AtBO\ being a bond disproportionated bismuthate. Despite the different arrangement of BiO$_6$ octahedra in \BBO\ and \AtBO , we observe a phonon mode in the Raman spectra of \AtBO\ at similar energy associated with breathing mode of the disproportionated \BBO .
Our results push forward the group of monovalent bismuthates $(M^{+1})_2$BiO$_3$ as a new class of bismuthates with novel topological properties. Doping of divalent atoms on the Ag-site and performing in-situ high-pressure experiments may allow us to destabilize the disproportionation in \AtBO, and achieve a metallic state and perhaps even superconductivity in this class of materials in analogy to \BBO . We envision that this work will open up the field of disproportionated oxides beyond those in the perovskite structure. 

\section{ACKNOWLEDGEMENTS}

We thank Frank Falkenberg, Christine Stefani, Sebastian Bette, and Armin Schulz for technical assistance with experiments.
We thank Tom Regier for help with XAS measurements. Research described in this paper was performed at the Canadian Light Source, which is supported by the Canada Foundation for Innovation, Natural Sciences and Engineering Research Council of Canada, the University of Saskatchewan, the Government of Saskatchewan, Western Economic Diversification Canada, the National Research Council Canada, and the Canadian Institutes of Health Research. We thank the Max Planck-UBC-UTokyo Center for Quantum Materials for support.

\bibliography{Ag2BiO3}

\renewcommand{\figurename}{Supplementary Figure}
\setcounter{figure}{0}

\renewcommand{\tablename}{Supplementary Table}
\setcounter{table}{0}

\pagebreak
\widetext
\begin{center}
\textbf{\large Electronic Structure of the Bond Disproportionated Bismuthate \AtBO - Supplemental Material}
\end{center}

\vspace{-0.8em}

\begin{figure}[h]
\includegraphics[width=16cm]{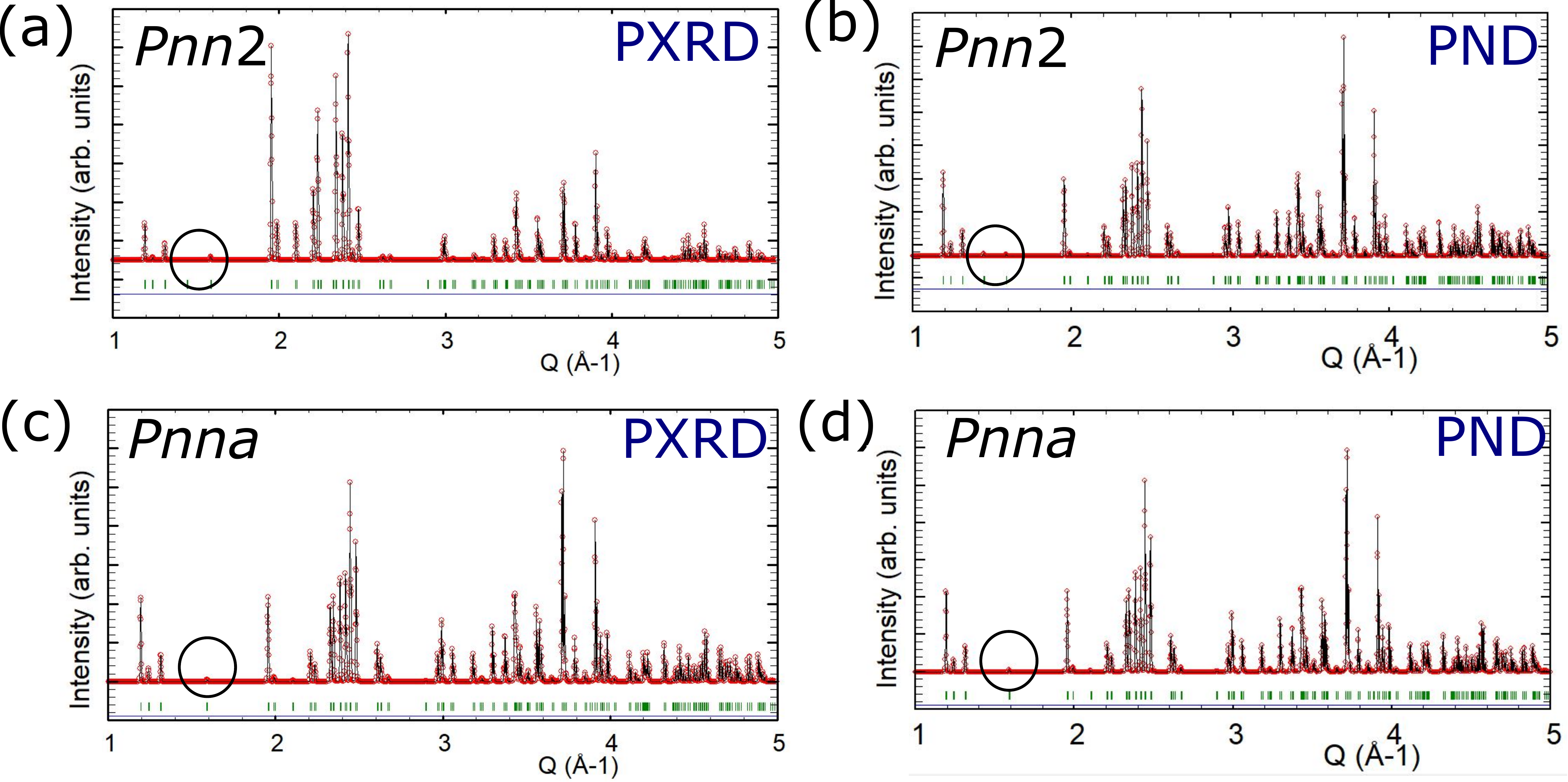}
\caption{Powder X-ray diffraction (PXRD) and powder neutron diffraction (PND) simulation Ag$_2$BiO$_3$ in the $Pnn2$ phase shown in (a) and (b), respecively. PXRD and PND simulation of Ag$_2$BiO$_3$ in the $Pnna$ phase shown in (c) and (d), respectively. Diffraction patterns are plotted as a function of the reciprocal lattice vector (Q). Black circles in the figures highlight one of the peaks only observable in the PND for the $Pnn2$ phase.}
\label{K_XPS}
\end{figure}

\begin{figure*}
\includegraphics[width=16cm]{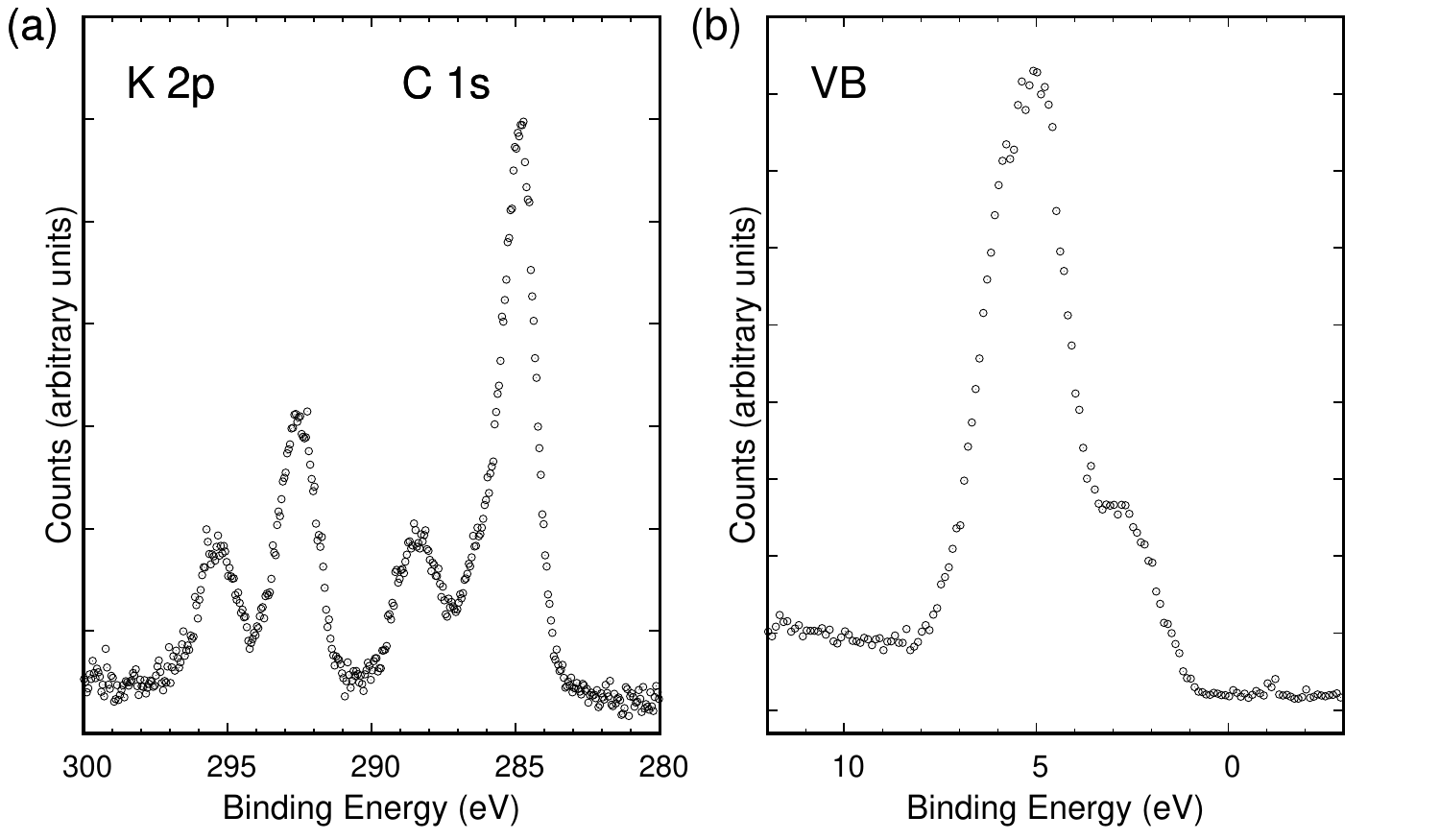}
\caption{X-ray photoelectron spectra of K $2p$ and C $1s$ (a) and of the valence band of Ag$_2$BiO$_3$ (b). In (b) the valence band edge is approx. 0.9 eV below the Fermi level.}
\label{K_XPS}
\end{figure*}

\begin{figure*}
\includegraphics[width=16cm]{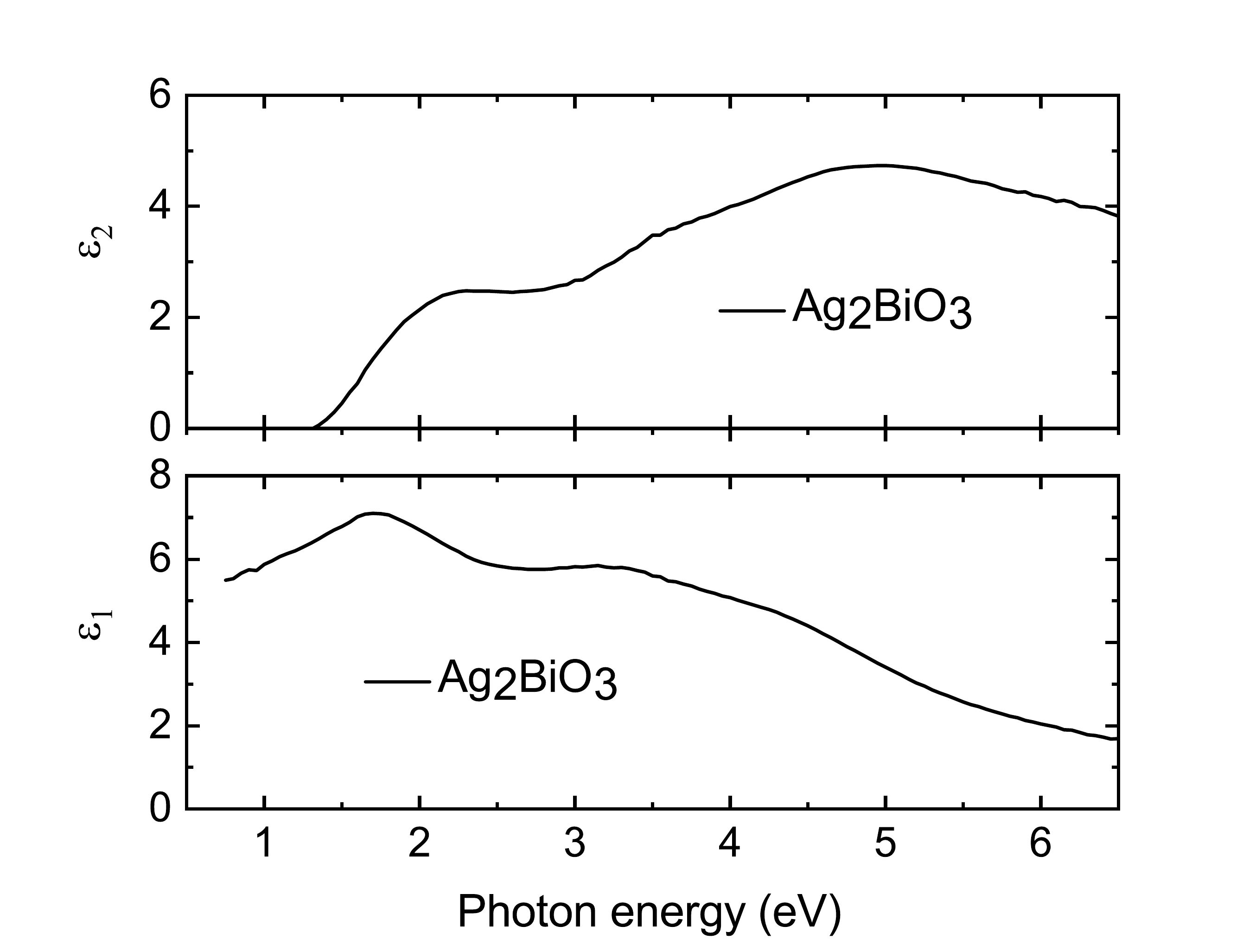}
\caption{Imaginary ($\epsilon_2$, top figure) and real ($\epsilon_1$, bottom figure) parts of the dielectric constant of \AtBO\ extracted from ellipsometry measurement.}
\label{Optical}
\end{figure*}

\begin{figure*}
\includegraphics[width=7cm]{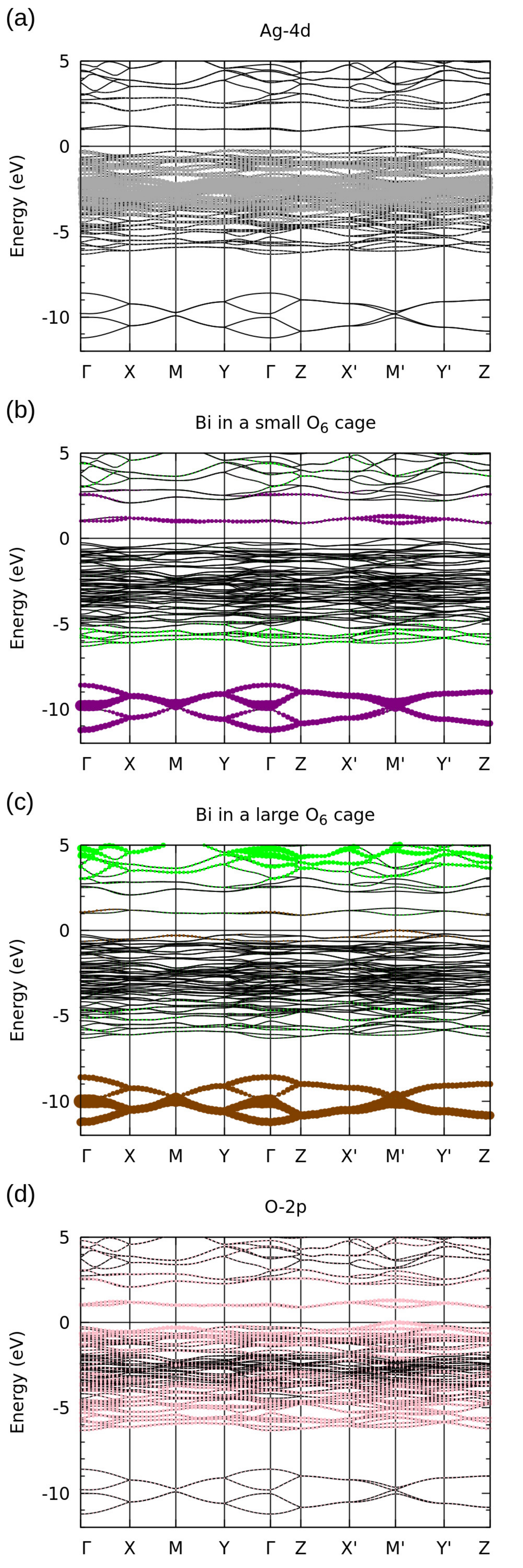}
\caption{(a) The DFT (LDA) band structures of \AtBO\ in the $Pnn2$ phase with fat band representation of atomic orbital contributions.
(a) The atomic orbital of the Ag-$4d$ shown in grey fat bands. (b) The fat bands of Bi-$s$ (green) and Bi-$p$ (purple) contribution in the small O$_6$ octahedral cage. (c) The fat bands of Bi-$s$ (green) and Bi-$p$ (brown) contribution in the large O$_6$ octahedral cage. (d) The atomic orbital of the O-$2p$ shown in pink fat bands.}
\label{FB}
\end{figure*}

\begin{table*}[]
\begin{ruledtabular}
\caption{Binding energy, assignment, full width at half-maximum (FWHM) and amount of the species observed via XPS. Please note that due to surface contaminations, error in the detector function and atomic sensitivity the amount might have an error in the range of 10-15\%.}
\label{SCTable}
\setlength{\extrarowheight}{12pt}
\begin{tabular}{|ccccc|}
\hline
element  & Assignment          & Binding energy {[}eV{]} & FWHM {[}eV{]} & Amount in {[}\%{]} \\
\hline
O $1s$     & Ag$_2$BiO$_3$    edge shared (peak C)    & 529.1                   & 1.1           & 11.9               \\
\hline
O $1s$     & \begin{tabular} {@{}c@{}}Ag$_2$BiO$_3$  corner shared (peak B) \\(small amounts of OH, carbonate contaminations) \end{tabular} & 530.8                   & 1.7           & 11.8               \\
\hline
O $1s$     & Water, hydroxides   (peak A)     & 532.6                   & 1.7           & 1.9                \\
\hline
C $1s$    & C-C, C-H   contaminations          & 284.8                   & 1.5           & 8.4                \\
\hline
C $1s$     & C-OH contaminations               & 285.7                   & 2.2           & 6.0                \\
\hline 
C $1s$     & carbonate                         & 288.5                   & 1.7           & 4.0                \\
\hline
K $2p_{3/2}$  & KOH                            & 292.6                   & 1.5           & 2.5                \\
\hline
Ag $3d_{5/2}$ & Ag$_2$BiO$_3$   (peak D2)      & 367.8                   & 1.2           & 16.1               \\
\hline
Bi $4f_{7/2}$ & Ag$_2$BiO$_3$   (peak G2)      & 158.0                   & 0.9           & 4.4                \\
\hline
Bi $4f_{7/2}$ & Ag$_2$BiO$_3$   (peak F2)      & 158.5                   & 1.4           & 5.2 \\
\hline           
\end{tabular}
\end{ruledtabular}
\end{table*}

\clearpage

\end{document}